\documentclass[a4paper,11pt]{article}
\usepackage{jheppub} % for details on the use of the package, please see the JINST-author-manual
\usepackage{lineno}
\usepackage{comment}
\usepackage[T1]{fontenc} % if needed
\usepackage{tikz}
\usepackage{amsmath,amssymb}
\usepackage{relsize}
\usepackage{latexsym}
\usepackage{marvosym}
\usepackage{leftidx}
\usepackage{xcolor}
\usepackage{diagbox}
\usepackage[T1]{fontenc}
\usepackage{array}
\usepackage{makecell}
\usepackage{csquotes}
\usepackage{enumitem}
\usepackage{setspace}
\usepackage{multirow}
\usepackage{subfig}
\usepackage{graphicx}
\usepackage{xcolor}
\usepackage{csquotes}
\usetikzlibrary{decorations.markings}
\usetikzlibrary{decorations.pathmorphing}
\usetikzlibrary{decorations.markings}
\usetikzlibrary{decorations.pathmorphing}
\title{\boldmath One-loop determinant in the extremal black hole from quasinormal modes}

\author{Jyotirmoy Mukherjee}
\affiliation[a]{ Department of Theoretical Physics\\
Tata Institute for Fundamental Research, Mumbai 400005, India.}
\emailAdd{jyotirmoy.mukherjee\_119@tifr.res.in}

\abstract{In this paper, we evaluate the one-loop partition function of a scalar field in the near-horizon geometry of the extremal Reissner Nordstr\"{o}m  black hole from an infinite product over quasinormal modes using the Denef-Hartnoll-Sachdev (DHS) formula. We show that the logarithmic divergent term of the one-loop partition function computed using the DHS formula agrees with the heat kernel method. Using the same formula, we also evaluate the one-loop partition function of a scalar field in the near-extremal Kerr-Newman black hole and observe that it reduces to the same in the near-horizon $AdS_2\times S^2$ geometry of the extremal Reissner Nordstr\"{o}m  black hole when the angular velocity at the horizon is tuned to $2\pi T_{BH}$ value. We observe that, for higher spin fields, the mode functions are not smooth at the horizon for certain quasinormal frequencies; therefore, we remove them to obtain the one-loop determinant.}

\begin{document}
\maketitle
\flushbottom
\section{Introduction}
The one-loop partition function of quantum fields is an important ingredient in evaluating the quantum corrections of black hole entropy. It has also been useful in extracting anomaly coefficients in even dimensions and $F$-termsin odd dimensions \cite{Giombi:2015haa, Fei:2015oha} in quantum field theory. 

Although the one-loop partition function is quite useful,  it is challenging to compute in an arbitrary curved background. The reason is that in most cases the spectrum of the kinetic operator is difficult to compute analytically. However, the heat-kernel method allows a useful tool in studying the divergences in one-loop determinants \cite{Vassilevich:2003xt}. In this method, one relates the coefficients of heat kernels with the gauge invariant geometric quantities, which are easily computable in a given background. Using the heat-kernel approach, the one-loop determinant of matter fields has been computed in non-extremal \cite{Sen:2012dw} and extremal black hole backgrounds \cite{Banerjee:2010qc,Sen:2012kpz}.

There has been a growing interest in studying extremal black holes because, in this case, one obtains a scale invariance near the horizon. Near the horizon of the extremal black holes,  a long throat is developed corresponding to $AdS_2$ geometry fibered over the angular directions. This is a place of physical interest because $AdS_2$ geometry is invariant under simultaneous rescaling of the time and radial directions, which is enhanced to the full conformal symmetry. There is another reason for interest in studying the extremal black hole, which is to understand its quantum nature.  To study the black hole as a quantum system, we require separating it from the rest of the environment, and the long throat of $AdS_2$ fibered over the angular direction provides an ideal space. Recently, the quantum corrections have been shown from the full higher-dimensional asymptotically flat or AdS geometry’s point of view and not focusing only near the throat \cite{Kolanowski:2024zrq}.

The quantum corrections to the near-extremal black hole entropies have been computed in \cite{Banerjee:2010qc,Sen:2012kpz} using a heat-kernel approach. However, there exists an independent way to evaluate Euclidean one-loop partition functions from infinite products over quasinormal modes, shown in \cite{Denef:2009kn} by Denef-Hartnoll-Sachdev. The derivation is based on the analytic properties of the partition function $\mathcal{Z}$ as a function of $m^2$. The quasinormal modes when Wick rotated correspond to the Euclidean mode at that particular value of $m^2$ which is regular at the horizon.

In this paper, we use the Denef-Hartnoll-Sachdev (DHS) prescription to compute the one-loop determinant of a scalar field in the near-horizon geometry of the extremal Reissner Nordstr\"{o}m black hole. The quasinormal modes are obtained by imposing the ingoing boundary condition at the horizon and the outgoing boundary condition at spatial infinity, since the geometry is asymptotically flat. The solutions are then matched in the overlapping region. From the discrete set of QNM in the full geometry, we obtain a nice integral expression of the one-loop partition function, which is given by
\begin{align}\label{1l}
 -\log\mathcal{Z}^{(1)}& =  \int_{\frac{a}{\epsilon}}^{\infty}\frac{dt}{2t}\frac{1+e^{-t}}{1-e^{-t}}\frac{e^{-t}+e^{-2t}}{(1-e^{-t})^3}
\end{align}
The coefficient of the logarithmic divergent piece can be obtained by expanding the integrand around $t=0$ and collecting the coefficient of $1/t$. The logarithmic divergent term for scalar field agrees with \cite{Banerjee:2010qc,Sen:2012kpz}.

We also study, the one-loop determinant of a scalar field \footnote{ one-loop determinant of higher spins in the extremal near-horizon AdS Reissner Nordstrom and AdS Kerr-Newman black holes has been computed in \cite{David:2021eoq}.}in the near-extremal Kerr-Newman black hole from the QNM spectrum. We show that, for a specific value of the angular velocity at the horizon, the one-loop partition function in the near-extremal (with a finite but small temperature) black hole reduces to the one-loop partition function in the near-horizon zero temperature extremal black hole. Explicitly, at $\Omega=2\pi T$, the one-loop partition function of the Kerr-Newman black hole at finite but small temperature reduces to \eqref{1l}. In \cite{Iliesiu:2021are,H:2023qko}, it is shown that at this value of the angular velocity at the horizon, the log of the index of black holes at finite temperature matches with the same
quantity computed using the near-horizon $AdS_2 \times S^2$ geometry at zero temperature \footnote{Note that, we compute the one-loop determinant from infinite product over QNM, which is in Lorentzian picture. Therefore, in the Euclidean case, the angular velocity will be $\Omega=-2\pi i T$.}. However, we can check the equality for the near-extremal black hole case. We also comment on the one-loop determinant of fields with spin $s\geq1$ where we observe that certain modes become irregular at the horizon and therefore have to be removed to obtain the partition function. Using this formalism, we also compute the one-loop partition function for the vector field at $\Omega=2\pi T$ and the logarithmic divergent piece which agrees with \cite{Sen:2012kpz}.

The organization of the paper is as  follows. We begin with the description of the near-horizon geometry of the extremal Reissner-Nordstr\"{o}m black hole which is the same as $AdS_2\times S^2$ spacetime. In section \eqref{sec3}, we directly compute one-loop regularized free energy of a massless scalar field in $AdS_2\times S^2$ from the eigenspectrum of the scalar field. In the next section, we briefly review the DHS formula which we use to evaluate the one-loop partition function in the extremal Reissner-Nordstr\"{o}m black hole. To obtain, the one-loop partition function we compute the quasinormal modes in the  extremal Reissner-Nordstr\"{o}m and Kerr-Newman black hole in section \eqref{RN} and \eqref{sec7} respectively. From the discrete set of QNM, we obtain the one-loop determinant of the scalar field in section \eqref{RN1} and \eqref{KN}. We also evaluate the one-loop partition function of a vector field in the near-extremal Kerr-Newman black hole from QNM in section \eqref{sec8}.
\section{Near-horizon geometry of the extremal Reissner-Nordstr\"{o}m black hole}
The Reissner Nordstr\"{o}m solution is specified by two parameters, mass of the black hole $M$ and the charge $Q$.

The metric is given by
\begin{align}\label{RN1}
& d s^2=-(1-\frac{r_+} { r})(1-\frac{r_-} { r}) d \tau^2+\frac{d r^2}{(1-\frac{r_+}{r})(1-\frac{r_-}{r})}+r^2\left(d \theta^2+\sin ^2 \theta d \phi^2\right),
\end{align}
where $r_+$ and $r_-$ are the locations of the outer and inner horizons are given by 
$$
r_{ \pm}=M \pm\left(M^2-Q^2\right)^{1 / 2}.
$$ The extremal limit corresponds to 
$$
r_+=r_-=M .
$$

Let us now define
\begin{align}
    t=\lambda \frac{\tau }{ r_+^2}, \quad \rho=\frac{(r-r_+)}{\lambda},
\end{align}
where $\lambda$ is a constant. We write the extremal solution in the new coordinate system and obtain

\begin{align}
d s^2 & =-\frac{\rho^2 r_+^4}{(r_++\lambda \rho)^2} d t^2+\frac{(r_++\lambda \rho)^2}{\rho^2} d \rho^2+(r_++\lambda \rho)^2\left(d \theta^2+\sin ^2 \theta d \phi^2\right).
\end{align}
The near-horizon limit can be taken by considering $\lambda \rightarrow 0$ limit. In this limit, one obtains
\begin{align}\label{nh1}
& d s^2=r_+^2\left(-\rho^2 d t^2+\frac{d \rho^2}{\rho^2}\right)+r_+^2\left(d \theta^2+\sin ^2 \theta d \phi^2\right).
\end{align}
Note that, in the limit, $\lambda \rightarrow 0$ but with fixed $\rho$ corresponds to the original coordinate $r \rightarrow r_+$. Therefore, the metric \eqref{nh1} describes the near-horizon geometry. It is also clear that the near-horizon geometry splits into the product of two spaces $AdS_2\times S^2$. Hence, the one-loop partition function of matter fields in the near-horizon geometry can be directly computed from the partition function on $AdS_2\times S^2$.
\section{One-loop determinant of a scalar field in $AdS_2\times S^2$}\label{sec3}

 In this section, we compute the regularized one-loop determinant of a massless real scalar on $AdS_2\times S^2$. We evaluate the free energy explicitly and express it as an integral form. We will match this integral with the free energy evaluated using the Denef-Hartnoll-Sachdev prescription. Therefore, this computation not only provides a test for the logarithmic divergent piece but also a consistency check for the entire partition function. 

 Here we would like to mention that, we compute regularized free energy by which we mean that we have incorporated the regularized volume of $AdS$ in the expression of the partition function. Therefore, the log divergent piece of this regularized partition function should be compared with the entropy correction obtained from the heat-kernel and regularization of the volume of $AdS$. To be more explicit, we compute 
 \begin{align}
       \log \mathcal{Z}[AdS_{{2}}\times S^2]^{(1)}&=\frac{\rm{Vol}(AdS_{2})}{4\pi}\,\sum_{n=0}^{\infty}g^{(0)}_n\int_{-\infty}^{\infty}d\lambda\,\mu^{(0)}_2(\lambda)\log\left(\lambda^2+(n+\frac{1}{2})^2\right).
 \end{align}
 Here $g_n^{(0)}$ is the degeneracy of a scalar Laplacian on $S^2$.
 The regularized volume of $AdS_{d+1}$ is computed using a unit ball realization prescription and is given by \cite{Diaz:2007an, Sun:2020ame}
\begin{equation}
\text{Vol}(\text{AdS}_{2}) = \pi^{\frac{1}{2}}\Gamma(-\frac{1}{2}).
\end{equation}
$\mu_2^{(0)}(\lambda)$ is the Plancherel measure for scalar field in $AdS_2$.
\begin{align}
    \mu_2^{(0)}(\lambda)=\lambda\tanh (\pi\lambda).
\end{align}
Note that, the regularized free energy contains the regularized volume of $AdS_2$ and the sum over eigenmodes on $S^2$ as well as $AdS_2$. Therefore, the log divergent piece of this regularized free energy should be matched against the entropy correction, because the entropy correction formula comes from the action which contains the regularized volume of $AdS$ and the effective matter Lagrangian \cite{Banerjee:2010qc}.

Since the eigenvalue and the degeneracy are known, we can compute the one-loop free energy following the logic of \cite{Banerjee:2010qc}.
\begin{align}\label{e1}
     \log \mathcal{Z}[AdS_{2}\times S^2]^{(1)}&=\frac{1}{4}\sum_{n=0}^{\infty}g^{(0)}_n\int_{-\infty}^{\infty}d\lambda\,\mu^{(0)}_2(\lambda)\log\left(\lambda^2+(n+\frac{1}{2})^2\right).
\end{align}
 
To obtain the expression \eqref{e1}, we use the eigenvalues of the scalar Laplacian on $AdS_2$ and $S^2$ \cite{Camporesi:1994ga}.
\begin{equation}\label{eigen}
\left( \Delta^{AdS_2}_{0}+\Delta_0^{S^2}\right) \psi^{\{\lambda, n\}} = -\left[  \lambda^2 + \frac{1}{4} +n(n+1)\right] 
 \psi_\lambda^{\{\lambda, n\}} ,
 \end{equation}
 The curvature induced mass  term $m_0^2$ on $ AdS_2\times S^2$ can be computed easily and it turns out to be zero.\footnote{The curvature induced mass term for scalar field on $AdS_b\times S^a$ is given by $ m_0^2=\frac{(a-b)(a+b-2)}{4}.$}
 
To perform the sum over eigenmodes, we use the integral representation of the logarithm.
\begin{align}\label{logiden}
    -\log \,y&=\int_0^{\infty}\frac{d\tau}{\tau}\left(e^{-y\tau}-e^{-\tau}\right)
\end{align}
\begin{align}
     \log \mathcal{Z}[AdS_{2}\times S^2]^{(0)}&=\frac{1}{4} \int_0^\infty \frac{d\tau}{\tau} \sum_{n=0}^\infty g^{(0)}_n
  \int_{-\infty}^{\infty} d\lambda 
  \mu_{0}^{(2)}(\lambda)  ( e^{ -\tau( \lambda^2 +(n+\frac{1}{2})^2 ) } - e^{-\tau} ).
\end{align}
 One can show that the second term where we perform the sum over degeneracy vanishes under the dimensional regularization prescription and we proceed with the first term to obtain
\begin{align}
  \log \mathcal{Z}[AdS_2\times S^2]^{(1)}&=-\int_0^{\infty}\frac{d\tau}{4\tau}e^{-\frac{\epsilon^2}{4\tau}}\sum_{n=0}^{\infty}g^{(0)}_n 
 \int_{-\infty}^{\infty} d\lambda 
  \mu^{(2)}_0 (\lambda)  \left(  e^{ - \tau (\lambda^2+(n+\frac{1}{2})^2}  \right)  .  
\end{align}
We have used $e^{-\frac{\epsilon^2}{4\tau}}$ factor to track the location of the branch cut. At the end of the computation, we will take $\epsilon \rightarrow 0$.
Let us now perform the integral over $\lambda$ by using the Hubbard-Stratonovich  trick
\begin{align}
     \log \mathcal{Z}[AdS_2\times S^2]^{(1)}&=\frac{-1}{4}\int_{-\infty}^{\infty}d\tau\int_{-\infty}^{\infty}\frac{du}{\sqrt{4\pi\tau^3}} \sum_{n=0}^{\infty}g^{(0)}_ne^{-\frac{\epsilon^2+u^2}{4\tau}}e^{-\tau(n+\frac{1}{2})^2}\int_{-\infty}^{\infty} d\lambda e^{i\lambda u} \mu_0^{(2)}(\lambda)\nonumber\\
     &=\frac{-1}{4}\int_{-\infty}^{\infty}d\tau\int_{-\infty}^{\infty}\frac{du}{\sqrt{4\pi\tau^3}} \sum_{n=0}^{\infty}g^{(0)}_ne^{-\frac{\epsilon^2+u^2}{4\tau}}e^{-\tau(n+\frac{1}{2})^2}W_0^{(2)}(u)
\end{align}
In \cite{Banerjee:2010qc}, the $\lambda$-integral was evaluated as a contour integral, running from $\infty$ to $0$ just below the real axis and returning to $\infty$ just above it. In our analysis, we find it more convenient to employ the Hubbard–Stratonovich trick, which allows us to perform the integral in terms of the Fourier transform of the Plancherel measure, denoted by $W_0^{(2)}$. The Fourier transform of scalar field in $AdS_2$ is given by \cite{Sun:2020ame, Mukherjee:2021alj}.
\begin{equation}\label{PlFourier}
W_0^{(2)}( u) = \frac{1+ e^{-u} }{ 1- e^{-u}} \frac{ e^{-\frac{1}{2} u} }{ ( 1- e^{-u} )} .
\end{equation}

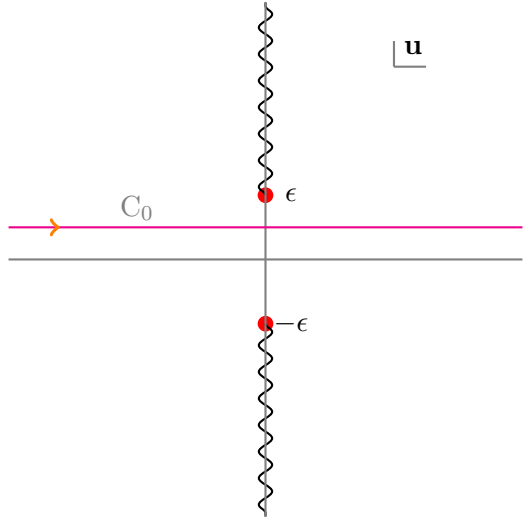
\begin{figure}[h]
\centering
\begin{tikzpicture}[thick,scale=0.85]
\filldraw[red] 
                (0,1) circle[radius=3pt]
                (0,-1) circle[radius=3pt];
\draw [decorate,decoration=snake] (0,-1) -- (0,-4);
\draw [decorate,decoration=snake] (0,1) -- (0,4);
%\draw (7.9,0) node {$\Re(w)$};
%\draw (0,6.5) node {$\Im(w)$};
%\draw (5.5,2.5) node {$\Re(w)>0$};
%\draw (-5.5,2.5) node{$\Re(w)<0$};
%\draw (3,-3) node{$|w|=1$};
\draw[gray,thick] (-2,0.8) node{$\mathbf{\rm{C_{0}}}$};
\draw (0.4,1) node{$\mathbf{\epsilon}$};
\draw (0.4,-1) node{$\mathbf{-\epsilon}$};
\draw[gray, thick] (0,0) -- (0,4);
\draw[gray, thick] (0,0) -- (0,-4);
\draw[gray, thick] (0,0) -- (4,0);
\draw[gray, thick] (0,0) -- (-4,0);
\draw
[
postaction={decorate,decoration={markings , 
mark=at position 0.20 with {\arrow[orange,line width=0.5mm]{>};}}}
][magenta, thick] (-4,0.5)--(0,0.5);
\draw[magenta, thick] (0,0.5)--(4,0.5);
\draw[gray, thick] (2,3) -- (2,3.4);
\draw[gray, thick] (2,3) -- (2.5,3);
\draw (2.3,3.3) node{$\mathbf{u}$};
\end{tikzpicture}
\caption{The $\rm{C_{0}}$ contour in the $u$-plane for  $AdS_{2}$} \label{fig4}
\end{figure}
 
 In $AdS_2$, the contour is shown in figure [\ref{fig4}] .
We  now perform the integral over $\tau$ and obtain
\begin{align}\label{KKsum}
        \log \mathcal{Z}[AdS_2\times S^2]^{(1)}&=-\int_{\rm{C_{0}}}\frac{du}{4\sqrt{u^2+\epsilon ^2}}\sum_{n=0}^{\infty}g^{(0)}_ne^{  \left(-(n+\frac{1}{2})\sqrt{u^2+\epsilon ^2}\right)}W_0^{(2)}(u).
\end{align}
Note that, after performing the integral over $\tau$, we can easily sum over the eigenmodes of $S^2$ in the exponent. Initially, we had a quadratic function of $n$ in the exponential which is difficult to sum analytically. But using the Hubbard-Stratonovich trick, the quadratic function transforms into a linear function which is now easy to sum. Let us now sum over eigenmodes on $S^2$ and take $\epsilon \rightarrow 0$, to obtain
\begin{align}\label{uintegralads}
    \log \mathcal{Z}[AdS_2\times S^2]^{(1)}&=-\frac{1}{2}\int_{-\infty}^{\infty}\frac{du}{2u}\frac{1+ e^{-u} }{ 1- e^{-u}}\frac{e^{-u(\frac{1}{2})}+e^{-u(\frac{3}{2})}}{(1-e^{-u})^2}\frac{ e^{-\frac{1}{2} u} }{ ( 1- e^{-u} )} \nonumber\\
      &=-\frac{1}{2}\int_{-\infty}^{\infty}\frac{du}{2u}\frac{1+ e^{-u} }{ 1- e^{-u}}\frac{e^{-u}+e^{-2 u }}{(1-e^{-u})^{3}}.
\end{align}
Since the integrand is invariant under $u\rightarrow -u$, we change the limit of integration from $0$ to $\infty$. Therefore, we write the partition function as
\begin{align}\label{1sc}
   \log \mathcal{Z}[AdS_2\times S^2]^{(1)}&= -\int_0^{\infty} \frac{du}{2u}\frac{1+ e^{-u} }{ 1- e^{-u}}\frac{e^{-u}+e^{-2 u }}{(1-e^{-u})^{3}}.
\end{align}

To collect the logarithmic divergent piece, we put a dimensionless cut-off to the lower limit of the integral. The lower limit is a ratio of the IR and UV cut-off.
\begin{align}\label{f2}
    \log \mathcal{Z}[AdS_2\times S^2]^{(1)}&= -\int_{\frac{a}{\epsilon}}^{\infty} \frac{du}{2u}\frac{1+ e^{-u} }{ 1- e^{-u}}\frac{e^{-u}+e^{-2 u }}{(1-e^{-u})^{3}}.
\end{align}
Therefore, the log of the regularized partition function can be obtained by expanding the integrand around $u=0$ and collecting the coefficient of $1/u$.
\begin{align}\label{f3}
  \mathcal{F}[AdS_2\times S^2]^{(1)}|_{\rm{log div}}&=- \log \mathcal{Z}[AdS_2\times S^2]^{(1)}|_{\rm{log div}}\nonumber\\
  &=-\frac{1}{180}\log \frac{a^2}{\epsilon}.
\end{align}
$\mathcal{F}[AdS_2\times S^2]^{(1)}|_{\rm{log div}}$ denotes the logarithmic divergent term of the regularised free energy of a massless scalar on $AdS_2\times S^2$. This logarithmic divergent piece agrees with \cite{Banerjee:2010qc}. Our objective is to compute the same using the Denef-Hartnoll-Sachdev prescription. 
\section{A brief review of the Denef-Hartnoll-Sachdev prescription}\label{sec4}
In this section, we briefly review the DHS prescription \cite{Denef:2009kn}. 
The one-loop determinant of a complex scalar field $\psi$ with mass $m$ is given by
\begin{align}
    \mathcal{Z}^{(1)}(m^2)&=\int D\psi e^{-\int \sqrt{g}d^d x \psi^* (-\nabla^2+m^2)\psi}\nonumber\\
    &\propto \frac{1}{(\det(-\nabla^2+m^2)},
\end{align}
where $\nabla^2$ is the kinetic operator in an arbitrary curved background.
In general, this one-loop determinant in a black hole background is a complicated object to compute. But Denef-Hartnoll-Sachdev prescribes the computation of this one-loop determinant of a scalar field in terms of infinite product over quasinormal modes. The prescription goes as follows.

For a non-compact background, one chooses the boundary conditions on fields at infinity and a mass parameter $\Delta (m^2)$. In $AdS$, $\Delta$ is the conformal dimension of the operator in the dual CFT. In the next step, we assume $\mathcal{Z}^{(1)}$ as  a meromorphic function in $\Delta(m^2)$ and analytically continue to the complex $\Delta$-plane. This meromorphic function is completely determined up to an entire function  from the location of its poles. This occurs when we have a zero mode of the wave equation of the corresponding field. This should also obey a periodicity condition in the Euclidean time direction. The positive frequency part of the solutions when Wick rotated satisfied the ingoing boundary condition at the horizon but the Wick rotated negative frequency part of the solutions satisfied the outgoing boundary condition. Putting the positive, negative, and the zero mode part of the solutions together, the one-loop determinant is expressed as
\begin{align}\label{DHS}
\mathcal{Z}^{(1)} =e^{\rm{Pol(\Delta)}} \prod_{\omega_*, \bar{\omega_*}} \prod_{p\geq 0} \left( p + \frac{i\omega_{*}}{2\pi T} \right)^{-1} \left( p - \frac{i \bar{\omega_*}}{2\pi T} \right)^{-1} .
\end{align}

Here $\bar{\omega}_*$ are the anti-quasinormal modes. $\rm{Pol(\Delta)}$ is the polynomial in $\Delta$ which is fixed by appropriate large $\Delta$ behavior.
\section{Perturbation equation in Reissner Nordstr\"{o}m  black hole }\label{RN}
The Reissner Nordstr\"{o}m solution is given in \eqref{RN1} and the inner and outer horizon radii in this solution are given by,
$$
r_{ \pm}=M \pm\left(M^2-Q^2\right)^{1 / 2}.
$$
Our goal is to obtain the one-loop determinant of a scalar field in the near-horizon and in the extremal limit of the RN black hole. The analytic expression of QNM in this background is difficult to obtain, however in \cite{Hod:2010hw}, it was shown that the wave equation for the massive scalar field can be solved analytically in the asymptotic region and in the near-horizon region.  The quasinormal modes are obtained by imposing the ingoing boundary condition at the horizon and the outgoing boundary condition at the spatial infinity. 

 We first write the ansatz for the modes
\begin{align}
    \Phi=e^{-i \omega t}e^{i m \phi}f(r)S_{l,m}(\theta).
\end{align}
Under this mode decomposition, the radial wave equation becomes \cite{Hod:2010hw}.
\begin{align}
    \frac{d}{dr} \left[ \Delta_H \frac{df(r)}{dr} \right] +  \left[ \frac{K_0^2}{\Delta_H}   - \mu^2r^2-\lambda  \right] f(r)= 0,
\end{align}
Here $\Delta_H=r^2-2M r+Q^2$ , $K_0=r^2 \omega$  and $ \lambda=l(l+1)$.
The angular part of the wave equation is also given by
\begin{align}
    \frac{d}{du} \left[ (1 - u^2) \frac{d S_{lm}}{du} \right] + \left[ + \lambda- \frac{(m )^2}{1 - u^2} \right] S_{lm} = 0, \quad u=\cos\theta
\end{align}

To obtain QNM, one imposes the ingoing boundary condition at the horizon and the outgoing boundary condition at the spatial infinity. 
\begin{align}
    f(r)&\sim e^{\frac{1}{2}\sqrt{\omega^2-\mu^2}r_*},\quad\quad r\rightarrow \infty\nonumber\\
    &\sim e^{-i \omega r_* },\quad\quad r\rightarrow r_+
\end{align}
  Here $r_*$ is the tortoise coordinate and $dr_*=\frac{r^2 dr}{\Delta}$.
Let us now define a new set of variables
\begin{align}\label{defz}
z &= \frac{r-r_{+}}{r_{+}}\nonumber\\
 \tau &= \frac{r_{+}-r_{-}}{r_{+}} \nonumber\\
 \hat{\omega} &=\frac{\omega}{2 \pi T}\nonumber\\
 k& = 2 \omega r_{+}.
\end{align}
With this new set of variables, the radial wave equation can be written as
\begin{align}\label{w1}
z(z+\tau) \frac{d^2 f(z)}{d z^2}+(2 z+\tau) \frac{d f(z)}{d z}+V(z) f(z)=0,
\end{align}
The potential $V(z)$ is given by
\begin{align}
V (z)&=\frac{ K^2} { r_{+}^2 z(z+\tau)}-\left[\mu^2 r_{+}^2(z+1)^2+l(l+1)\right],\end{align}
where $K$ is defined in the following form
\begin{align}\label{KV}
K&=r_{+}^2 \omega z^2+r_{+} k z+r_{+} \hat{\omega} \frac{\tau} { 2}.
\end{align}

It is not very easy to obtain the solution of the wave equation \eqref{w1} for the entire geometry. However, we can obtain the solution exactly for the asymptotic region ($r\rightarrow \infty$) and near the horizon ($r\rightarrow r_+$). We then match the two solutions in the overlapping region $\tau\ll z\ll 1$, where $z$ and $\tau$ are defined in \eqref{defz}. In this way, we can find the solution for the outside of the horizon to asymptotic infinity.
\paragraph{Solution in the asymptotic region}
Let us first consider the wave equation in the asymptotic region $r\rightarrow \infty$ or $z\gg \tau$ with $\omega M \ll 1$. In this regime, the wave equation \eqref{w1} can be approximated as
\begin{align}
z^2 \frac{d^2 f(z)}{d z^2}+2 z \frac{d f(z)}{d z}+V_{\text {asymp }} f(z)=0,
\end{align}
The potential in the asymptotic regime is given by
\begin{align}
  V_{\text {asymp }}  &=\left(\omega^2-\mu^2\right) r_{+}^2 z^2+2\left(\omega k-\mu^2 r_{+}\right) r_{+} z-\left[\mu^2 r_{+}^2+l(l+1)-k^2\right].
\end{align}
To obtain $V_{\text {asymp }}$, we focus on the asymptotic regime, which is $z\gg \tau$. Therefore, the first two terms of $K$  have a dominant contribution over the last term in the potential $V_{\text {asymp }}$.
In this asymptotic regime, the wave equation can be solved analytically in terms of the confluent hypergeometric equation. The solution which satisfies the outgoing boundary condition is given by
\begin{align}\label{sa}
f(z)&=C_1\left(2 i (\omega^2-\mu^2)^{\frac{1}{2}} r_{+}\right)^{\frac{1}{2}+i \delta} z^{-\frac{1}{2}+i \delta} e^{-i \sqrt{\omega^2-\mu^2} r_{+} z} F\left(\frac{1}{2}+i \delta+i \kappa, 1+2 i \delta, 2 i \sqrt{\omega^2-\mu^2} r_{+} z\right)\nonumber\\
&+C_2\left(2 i (\omega^2-\mu^2)^{\frac{1}{2}} r_{+}\right)^{\frac{1}{2}-i \delta} z^{-\frac{1}{2}-i \delta} e^{-i \sqrt{\omega^2-\mu^2} r_{+} z} F\left(\frac{1}{2}-i \delta+i \kappa, 1-2 i \delta, 2 i \sqrt{\omega^2-\mu^2} r_{+} z\right)
\end{align}
$\delta$ and $\kappa$ are given by
\begin{align}\label{del}
    \delta^2=k^2-\mu^2 r_{+}^2-(l+\frac{1}{2})^2, \quad \kappa=\frac{\omega k-\mu^2 r_{+}}{\sqrt{\omega^2-\mu^2}}. 
\end{align}
Here, $F(a,b,z)$ denotes the confluent hypergeometric function and $C_1$, $C_2$ are two constants that will be determined from the matching condition in the overlapping region.
\paragraph{Solution near the horizon}
The radial equation can also be solved analytically, in the near-horizon regime $r\rightarrow r_{+}$ or $z \ll 1$.
The wave equation is written as 
\begin{align}\label{wv}
    z(z+\tau) \frac{d^2 f(z)}{d z^2}+(2 z+\tau) \frac{d f(z)}{d z}+V(z) f(z)=0,
\end{align}
where the near-horizon potential $V_{nh}$ can be approximated as 
\begin{align}
    V_{nh}&=-\Big[\mu^2 r_{+}^2+l(l+1)\Big]+\frac{(k z+\hat{\omega} \frac{\tau}{2} )^2 }{z(z+\tau)},\quad\quad z\ll 1.
\end{align}
To obtain, the near-horizon potential $V_{nh}$, we keep the last two terms of $K$ which has a dominant contribution over the first term  under the condition $r_+ T\ll 1$, where $T$ is the temperature of the black hole.\footnote{Since in the near-horizon-extremal regime, $\frac{z^2}{\tau}\ll 1$ and $r_+T\ll 1$, we obtain $\frac{z^2}{\tau}\ll 1 \ll \frac{1}{r_+ T}$. This condition allows us to drop the first term in $K$ of the potential in the near-horizon regime. }

In the low-frequency, near-horizon region, one obtains the solution which satisfies the ingoing boundary condition. The physical solution is given by
\begin{align}\label{sn}
f(z)=z^{-\frac{i}{2} \hat{\omega}}\left(\frac{z}{\tau}+1\right){ }^{2i\left(\frac{1}{2} \hat{\omega}-k\right)} {}_2F_1\left(\frac{1}{2}+i \delta-i k, \frac{1}{2}-i \delta-i k ; 1-i \hat{\omega} ;-z / \tau\right),
\end{align}

\paragraph{Solution in the overlapping region} To obtain a unique solution which satisfies the ingoing boundary condition at the horizon and outgoing boundary condition at the asymptotic infinity \footnote{We solve the wave equations in the RN black hole in flat space, the boundary condition in the asymptotic infinity should be outgoing.},
we require matching the solutions in the region of overlap $\tau\ll z\ll 1$.
In the $z\ll 1$,the solution \eqref{sa} becomes,
\begin{align}
f(z) \sim C_1\left(2 i \sqrt{\omega^2-\mu^2} r_{+}\right)^{\frac{1}{2}+i \delta} z^{-\frac{1}{2}+i \delta}+C_2(\delta \rightarrow-\delta) .
\end{align}

The $z \gg \tau$ limit , the solution \eqref{sn}
becomes
\begin{align}\label{s1}
 f(z)\sim   \tau^{\frac{1}{2}-i \delta-i \hat{\omega} / 2} \frac{\Gamma(2 i \delta) \Gamma(1-i \hat{\omega})}{\Gamma\left(\frac{1}{2}+i \delta-i k\right) \Gamma\left(\frac{1}{2}+i \delta-i \hat{\omega}+i k\right)} z^{-\frac{1}{2}+i \delta}+(\delta \rightarrow-\delta)
\end{align}
Note that, the solution exhibits the power law nature in this region.

We now match these solutions to obtain the coefficients $C_1$ and $C_2$.
\begin{align}
& C_1=\tau^{\frac{1}{2}-i \delta-i \hat{\omega} / 2} \frac{\Gamma(2 i \delta) \Gamma(1-i \hat{\omega})}{\Gamma\left(\frac{1}{2}+i \delta-i k\right) \Gamma\left(\frac{1}{2}+i \delta-i \hat{\omega}+i k\right)}\left(2 i \sqrt{\omega^2-\mu^2} r_{+}\right)^{-\frac{1}{2}-i \delta}, \\
& C_2=\tau^{\frac{1}{2}+i \delta-i \hat{\omega} / 2} \frac{\Gamma(-2 i \delta) \Gamma(1-i \hat{\omega})}{\Gamma\left(\frac{1}{2}-i \delta-i k\right) \Gamma\left(\frac{1}{2}-i \delta-i \hat{\omega}+i k\right)}\left(2 i \sqrt{\omega^2-\mu^2} r_{+}\right)^{-\frac{1}{2}+i \delta} .
\end{align}

\begin{comment}We now take the solution \eqref{sa} in the $z\rightarrow \infty$ regime,
\begin{align}
f(z) \sim & {\left[C_1\left(2 i \sqrt{\omega^2-\mu^2} r_{+}\right)^{i \kappa} \frac{\Gamma(1+2 i \delta)}{\Gamma\left(\frac{1}{2}+i \delta+i \kappa\right)} z^{-1+i \kappa}+C_2(\delta \rightarrow-\delta)\right] e^{i \sqrt{\omega^2-\mu^2} r_{+} z} } \\
& +\left[C_1\left(2 i \sqrt{\omega^2-\mu^2} r_{+}\right)^{-i \kappa} \frac{\Gamma(1+2 i \delta)}{\Gamma\left(\frac{1}{2}+i \delta-i \kappa\right)} z^{-1-i \kappa}(-1)^{-\frac{1}{2}-i \delta-i \kappa}+C_2(\delta \rightarrow-\delta)\right] e^{-i \sqrt{\omega^2-\mu^2} r_{+} z}
\end{align}
\end{comment}
%\begin{comment}
%The solutions in the asymptotic region \eqref{sa} and the near-horizon \eqref{sn} can be matched in the overlapping region $\tau\ll z\ll 1$ to determine $C_1$ and $C_2$ within the low frequency regime $\omega r_{+}\ll 1.$
%\begin{align} C_1 & =\tau^{\frac{1}{2}-i \delta-i \hat{\omega} / 2} \frac{\Gamma(2 i \delta) \Gamma(1-i \hat{\omega})}{\Gamma\left(\frac{1}{2}+i \delta-i k\right) \Gamma\left(\frac{1}{2}+i \delta-i \hat{\omega}+i k\right)}\left(2 i \sqrt{\omega^2-\mu^2} r_{+}\right)^{-\frac{1}{2}-i \delta}, \\ C_2 & =\tau^{\frac{1}{2}+i \delta-i \hat{\omega} / 2} \frac{\Gamma(-2 i \delta) \Gamma(1-i \hat{\omega})}{\Gamma\left(\frac{1}{2}-i \delta-i k\right) \Gamma\left(\frac{1}{2}-i \delta-i \hat{\omega}+i k\right)}\left(2 i \sqrt{\omega^2-\mu^2} r_{+}\right)^{-\frac{1}{2}+i \delta} .\end{align}
%\end[comment}]
\paragraph{Consistency check} \footnote{We thank Shiraz Minwalla for pointing out this consistency check of the solution in the overlapping region. }Here we provide an important consistency check of the solution for the entire geometry. In the far regime, $z\gg \tau$, we found the dominant contribution of the first two terms of $K$ in the potential \eqref{KV}. Therefore, in the far regime ($z\gg \tau$), we write
\begin{align}
    K=r_{+}^2 \omega z^2+r_{+} k z+r_{+} \hat{\omega} \frac{\tau} { 2}\approx r_{+}^2 \omega z^2+r_{+} k z.
\end{align}In the near-horizon regime ($z\ll 1$), the last two terms dominated over the first term in the definition of $K$.
\begin{align}
   K=r_{+}^2 \omega z^2+r_{+} k z+r_{+} \hat{\omega} \frac{\tau} { 2}\approx r_{+} k z+r_{+} \hat{\omega} \frac{\tau} { 2}
\end{align}
Therefore, the solution in the overlapping region should be independent of both the first and third terms in the definition of $K$ (given in \eqref{KV}). This is simply because, in the overlapping regime, the solution is not governed by both terms. Now, if we solve the wave equation \eqref{wv}, keeping only the middle term of $K$ in the potential, we should obtain the power law behavior of the solution given in \eqref{s1}. Note that, the solution in \eqref{s1} in the overlapping region is obtained by matching the solutions in the asymptotic region and in the near-horizon region. Therefore, this method provides us to obtain the solution in the overlapping region in another way. We solve the wave equation keeping only the middle term of $K$ in the overlapping region and indeed obtain the same power law behavior of the solution given in \eqref{s1}. 

Since we obtain $C_1$ and $C_2$ from the matching condition, we can plug them in \eqref{sa} and take $z\rightarrow \infty$ limit and impose vanishing boundary condition to the ingoing modes which result in the following condition
\begin{align}
&\frac{\Gamma(2 i \delta) \Gamma(1+2 i \delta)\left(2 \tau \sqrt{\mu^2-\omega^2} r_{+}\right)^{-i \delta}}{\Gamma\left(\frac{1}{2}+i \delta-i \kappa\right) \Gamma\left(\frac{1}{2}+i \delta-i k\right) \Gamma\left(\frac{1}{2}+i \delta-i \hat{\omega}+i k\right)}\nonumber\\
&\quad\quad\quad\quad\quad+\frac{\Gamma(-2 i \delta) \Gamma(1-2 i \delta)\left(2  \tau \sqrt{\mu^2-\omega^2} r_{+}\right)^{i \delta}}{\Gamma\left(\frac{1}{2}-i \delta-i \kappa\right) \Gamma\left(\frac{1}{2}-i \delta-i k\right) \Gamma\left(\frac{1}{2}-i \delta-i \hat{\omega}+i k\right)}=0 .
\end{align}
 Let us now consider the low frequency regime $\omega r_+\ll \mu r_+\ll 1$, we have
\begin{align}\label{deltaapprox}
    \delta^2\approx-\mu^2r_+^2-(l+\frac{1}{2})^2.
\end{align}
Considering the branch, $\delta=i \sqrt{\mu^2r_+^2+(l+\frac{1}{2})^2}$, and in the near-extremal regime $\tau \ll 1$, one obtains
\begin{align}\label{ep}
 \epsilon=  \left(\tau \sqrt{\mu^2-\omega^2}r_{+}\right)^{-2i\delta}=\left(\tau \sqrt{\mu^2-\omega^2}r_{+}\right)^{2\sqrt{\mu^2r_+^2+(l+\frac{1}{2})^2}}\ll 1.
\end{align}
Note that, for $l\geq 0$ modes, the factor $2\sqrt{\mu^2r_+^2+(l+\frac{1}{2})^2}\geq 1$ and therefore, the above approximation is justified.
So we can write the quasinormal mode condition in the following way
\begin{align}\label{eq}
\frac{1}{\Gamma\left(\frac{1}{2}-i \delta-i \hat{\omega}+i k\right)}=\mathcal{F} \times\left(\tau \sqrt{\mu^2-\omega^2}r_{+}\right)^{2\sqrt{\mu^2r_+^2+(l+\frac{1}{2})^2}}=O(\epsilon),
\end{align}
where $\mathcal{F}$ is given by
\begin{align}\mathcal{F} &=\frac{[\Gamma(2 i \delta)]^2 \Gamma\left(\frac{1}{2}-i \delta-i \kappa\right) \Gamma\left(\frac{1}{2}-i \delta-i k\right)}{ [\Gamma(-2 i \delta)]^2 \Gamma\left(\frac{1}{2}+i \delta-i \kappa\right) \Gamma\left(\frac{1}{2}+i \delta-i k\right) \Gamma\left(\frac{1}{2}+i \delta-i \hat{\omega}+i k\right)}, \quad\quad\kappa =\frac{\omega k-\mu^2 r_+}{\sqrt{\omega^2-\mu^2}}.
\end{align}

Note that $\mathcal{F}$ has a nice behavior in the near-extremal limit. In this limit, the term on the right-hand side of equation \eqref{eq} is of order $O(\epsilon)$ where $\epsilon$ is given in \eqref{ep}.

Therefore, the quasinormal modes can be obtained from
\begin{align}
    \frac{1}{2}-i\delta-i\hat{\omega}+i k=-n.
\end{align}
We now use the condition given in \eqref{deltaapprox}, and find the quasinormal modes:
\begin{align}\label{QNMRN}
    \frac{\omega_*}{2\pi T}&\approx -\delta-i(n+\frac{1}{2})\nonumber\\
   & \approx -i\left(n+\frac{1}{2}+\sqrt{\mu^2r_+^2+(l+\frac{1}{2})^2}\right)\nonumber\\
   &=-i\left(n+\Delta_{AdS_2}\right),
\end{align}
where the quantity $\Delta_{AdS_2}$ is the conformal dimension of the effective 
field in the near-horizon $AdS_2$ region, given by the equation
\begin{align}
   \Delta_{AdS_2}
   &= \frac{1}{2}+\sqrt{\frac{1}{4}+m_{\text{eff}}^2\,r_+^2}\nonumber\\
  &= \frac{1}{2}+\sqrt{\frac{1}{4}+\mu^2 r_+^2 + l(l+1)}
\end{align}
Here $m_{\text{eff}}^2=\mu^2+\frac{l(l+1)}{r^2_+}$ is the effective mass squared of the field in $AdS_2$.  
% For a bulk scalar of bare mass $\mu$ propagating on the black hole background,  
% the Kaluza-Klein reduction on the $S^2$ introduces an additional contribution 
% to the mass through the spherical harmonics with angular momentum $l$.  
% Explicitly,
% \begin{align}
%    m_{\text{eff}}^2\,r_+^2 
%    &= \mu^2 r_+^2 + l(l+1).
% \end{align}
% Substituting this back into the expression for $\Delta_{AdS_2}$ gives
% \begin{align}
%    \Delta_{AdS_2}
%    &= \frac{1}{2}+\sqrt{\frac{1}{4}+\mu^2 r_+^2 + l(l+1)}.
% \end{align}
Thus, the quasinormal mode spectrum \eqref{QNMRN} naturally organizes itself 
in terms of the $AdS_2$ scaling dimension, with the contribution from the 
compact $S^2$ encoded in the KK tower through the $l(l+1)$ term.
 We will use the analytic expression  of the near-horizon extremal RN black hole given in \eqref{QNMRN} to evaluate the one-loop determinant of the scalar field using the DHS formula.
\subsection{One-loop determinant of a scalar field}\label{RN1}
In this section, we evaluate one-loop determinant of a scalar field in near-horizon extremal limit of RN black hole using the DHS prescription.

\begin{align}\label{one-loopdetsc}
    -\log\mathcal{Z}^{(1)}&=\sum_{p>0}\sum_{n,l\geq 0}\sum_{m}\log\left(|p|+i\frac{\omega_*}{2\pi T}\right)+\log\left(|p|-i\frac{\bar{\omega_*}}{2\pi T}\right)
    +\frac{1}{2}\sum_{n,l\geq0}\log\left(+i\frac{\omega_*}{2\pi T}\right)+\log\left(-i\frac{\bar{\omega_*}}{2\pi T}\right)\nonumber\\
    &=\sum_{p>0}\sum_{n,l\geq 0}(2l+1)\log\left((p+n+\frac{1}{2})^2+\delta^2\right)
    +\frac{1}{2}\sum_{n,l\geq0}(2l+1)\log\left(+(n+\frac{1}{2})^2+\delta^2\right)\nonumber\\
    &=+\frac{1}{2}\sum_{k',l=0}^{\infty}(2l+1)(2k'+1)\log\left((k'+\frac{1}{2})^2+\delta^2\right)
\end{align}
In the final line, we convert the sum over $n$ and $p$ into a single sum over $k'$ with appropriate multiplicity factor $(2k'+1)$ \footnote{The reader may wonder why the sum over $n$ is extended all the way to infinity. 
A natural cutoff for this sum would be at $
n_{\max} \sim \frac{1}{T r_+},
$
since the matching condition is derived under the approximation $\omega r_+ \ll 1$. 
However, as we demonstrate in Appendix~\ref{jna}, the contribution from terms with 
$
n \gtrsim \frac{1}{T r_+}
$
does not affect the logarithmically divergent part of the determinant.
}. For the massless perturbation, in the near-horizon extremal RN black hole,  the one-loop determinant becomes
\begin{align}
     -\log\mathcal{Z}^{(1)} &=+\frac{1}{2}\sum_{k',l=0}^{\infty}(2l+1)(2k'+1)\log\left((k'+\frac{1}{2})^2-\mu^2r_+^2-(l+\frac{1}{2})^2\right)
\end{align}
To write the one-loop partition function, we consider the modes $\omega r_+\ll 1$, and we show that with these modes one obtains the one-loop partition function of a scalar field in the near-horizon regime. 
To perform the sum over modes, we use the integral representation of logarithm.
\begin{align}\label{logiden}
    -\log \,y&=\int_0^{\infty}\frac{d\tau}{\tau}\left(e^{-y\tau}-e^{-\tau}\right)
\end{align}
Note that, the second term involves the sum which can be shown to be zero under the dimension regularization \cite{Mukherjee:2021alj}.

Proceeding with the first term
\begin{align}
  -\log\mathcal{Z}^{(1)}&=
   \int_{\epsilon}^{\infty} \frac{d\tau}{2\tau} e^{-\frac{\epsilon^2}{4\tau}}\sum_{k',l=0}^{\infty}(2l+1) (2k'+1) e^{-\tau \left( \left( k'+\frac{1}{2}\right)^2 +\nu^2\right)}, \quad \nu^2=-\mu^2r_+^2-(l+1/2 )^2.
\end{align}
We now use the Hubbard-Stratonovich trick to perform the sum over $k'$.
\begin{align}
    \sum_{k'=0}^{\infty} (2k+1) e^{-\tau \left( k'+\frac{1}{2}\right)^2}
= \int_{C}\frac{du}{\sqrt{4\pi \tau}} e^{-\frac{u^2}{4\tau}} f(u).
\end{align}
where $C=\mathbb{R}+i\delta$ and $f(u)$ is given by
\begin{align}
    f(u) = \sum_{k'=0}^{\infty} (2k'+1)e^{iu(k'+\frac{1}{2})}
= \frac{e^{\frac{i u}{2}} \left(1+e^{i u}\right)}{\left(-1+e^{i u}\right)^2}
\end{align}
Note that, using the Hubbard-Stratonovich trick, we can easily sum over $k'$.  Let us now perform the integral over $\tau$ 
\begin{align}
   -\log\mathcal{Z}^{(1)}&=\sum_{l\geq 0}(2l+1)\int_{C} \frac{du}{2\sqrt{u^2+\epsilon^2}}e^{- \nu \sqrt{u^2+\epsilon^2} }f(u)
\end{align}
This integral has a branch cut with the branch points $u=\pm i \epsilon$. To perform the integral, we wrap the contour along the branch cuts and rotate it by replacing $u=i t$. We isolate the convergent part of the sum over $l$ along the one side of the branch cut. At this stage, we can also take the massless limit $\mu\rightarrow 0$ and obtain
\begin{align}
    -\log\mathcal{Z}^{(1)}&=\int_{\epsilon}^{\infty}\frac{dt}{2\sqrt{t^2-\epsilon^2}}\sum_{l\geq 0} f(u=i t)e^{-t(l+\frac{1}{2})}
\end{align}
We perform the sum over $l$ and take $\epsilon \rightarrow 0$ limit, to obtain
\begin{align}\label{DHS1}
     -\log\mathcal{Z}^{(1)}&=\int_0^{\infty}\frac{dt}{2t}\frac{1+e^{-t}}{1-e^{-t}}\frac{e^{-t}+e^{-2t}}{(1-e^{-t})^3}
\end{align}
Note that, we obtain the same expression of the one-loop partition function for the scalar field in $AdS_2\times S^2$ in \eqref{1sc}. Therefore, we show that, the one-loop determinant of a scalar field in the near-horizon extremal RN black hole computed using the DHS prescription agrees with a direct evaluation of the partition function in $AdS_2\times S^2$. The logarithmic correction can also be obtained in a similar way where we put a dimensionless cutoff in the lower limit of the integral and expand the integrand around $t=0$ and collect the coefficient of $1/t$ of the integrand. Finally, we obtain the same number for the log divergent piece as given in \eqref{f3}.

It is important to note that the integral representation of the one-loop determinant obtained from the quasinormal modes in \eqref{DHS1} agrees precisely with the heat-kernel evaluation in \eqref{f2}. This establishes that $e^{\mathrm{Pol}(\Delta)}=1$ in the expression \eqref{DHS}.

\section{Kerr-Newman solution}
\label{KN}

In the following sections, we evaluate one-loop determinant of a scalar field in the near-extremal Kerr-Newman black hole using the DHS prescription. We show that, in the near-extremal limit the one-loop determinant at a finite but very small temperature black hole agrees with the one-loop determinant in the near-horizon extremal RN black hole at zero temperature when the angular velocity at the horizon is tuned to a specific value $\Omega=2\pi T$. The equality of the partition function of the finite temperature black hole and the near-horizon zero temperature black hole has been explicitly shown in \cite{Iliesiu:2021are,H:2023qko}. Here we can only show the equality of the one-loop partition functions between the near-extremal at a finite but very small temperature with the zero temperature extremal black hole when $\Omega\rightarrow 2\pi T$.

Let us begin with the Kerr-Newman solution.
The Kerr-Newman metric is characterized by three parameters, the black hole mass M, the charge Q and angular momentum per unit mass $a=\frac{J}{M}.$ In the $Q\rightarrow 0$ limit, it reduces to the rotating Kerr metric and in the $a\rightarrow 0$ limit, it reduces to the Reissner-Nordstr\"{o}m metric.

The Kerr-Newman solution is given by
\begin{align}
    ds^2 &= -\frac{r^2 + a^2 \cos^2 \theta - 2Mr + Q^2}{r^2 + a^2 \cos^2 \theta} dt^2 + \frac{r^2 + a^2 \cos^2 \theta}{r^2 + a^2 \cos^2 \theta - 2Mr + Q^2} dr^2 + (r^2 + a^2 \cos^2 \theta)d\theta^2\nonumber \\
    & \quad + \frac{(r^2 + a^2 \cos^2 \theta)(r^2 + a^2) + (2Mr - Q^2)a^2 \sin^2 \theta}{r^2 + a^2 \cos^2 \theta} \sin^2 \theta d\phi^2\nonumber \\
    & \quad + \frac{2(Q^2 - 2Mr)a}{r^2 + a^2 \cos^2 \theta} \sin^2 \theta dtd\phi,\\
     A_\mu dx^\mu &= \frac{-Qr}{r^2 + a^2 \cos^2 \theta} \left( dt - a \sin^2 \theta d\phi \right) + \frac{Q r_+}{r_+^2 + a^2} dt
\end{align}
The inner and outer horizons are located at
\begin{equation}
    r_\pm = M \pm \sqrt{M^2 - Q^2 - a^2}.
\end{equation}
The inverse temperature and the angular velocity at the horizon are given by
\begin{align}
    \beta&=4\pi\frac{(r_{+}^2+a^2)r_{+}}{r_{+}^2-a^2-Q^2}, \quad\quad \Omega=\frac{a^2}{r_{+}^2+a^2}.
\end{align}
We now define a set of following variables
\begin{equation}\label{deltabar}
    \rho^2 = r^2 + a^2 \cos^2 \theta, \quad \bar{\Delta}_H = r(r - 2M) + a^2+ Q^2.
\end{equation}
The metric under the following parametrization becomes
\begin{align}
    ds^2 &= -\frac{\rho^2 - 2M r+Q^2}{\rho^2} dt^2 - \frac{4 a M r \sin^2 \theta-Q^2 a^2 \sin^2\theta}{\rho^2} dtd\phi + \frac{\rho^2}{\bar{\Delta}_H+Q^2} dr^2 \\
    & \quad + \rho^2 d\theta^2 + \frac{(r^2 + a^2)^2 - a^2 \bar{\Delta}_H \sin^2 \theta-Q^2 a^2\sin^2\theta}{\rho^2} \sin^2 \theta d\phi^2,
\end{align}
To solve the wave equations in this background, we will follow the set of differential equations introduced by Dudley and Finley to describe the perturbations of the Kerr-Newman space-time \cite{Berti:2005eb}.

%step:
%1. Radial Teukolsky equation is given. Copy it
%Take it, then use extremal kerr ads2 case. use ads_2 times s2 paper, do the same.

%or near-horizon casradiale , remember this is a flat space, use Law's arguement. Will get the same answer.
%\end{comment}
\subsection{Dudley-Finley perturbation equations}\label{DF}Dudley and Finley reduced the perturbation equations of a fixed Kerr–Newman black hole to a pair of differential equations: one governing the angular part and the other the radial part~\cite{Dudley:1977zz,Berti:2005eb}.
 %These set of equations are the generalization of Teukolsky equations for Kerr black hole \cite{PhysRevLett.29.1114}.
We write the mode expansion using the following ansatz
\begin{align}
    \Phi(t,r,\theta,\phi)=e^{i m \phi}e^{-i \omega t}f(r) S_{\ell,m}(\theta)
\end{align}The radial part of the equation for the massless scalar is given by \cite{Berti:2005eb}
\begin{align}
    \frac{d}{dr} \left[ \bar{\Delta}_H \frac{df(r)}{dr} \right] +  \left[ \frac{\bar{K}_0^2}{\bar{\Delta}_H}   - \lambda  \right] f(r)= 0,
\end{align}
where $\bar{\Delta}_H$ is defined in \eqref{deltabar}.
The angular part of the wave equation is given by
\begin{align}
    \frac{d}{du} \left[ (1 - u^2) \frac{d S_{lm}}{du} \right] + \left[ (\omega u)^2 + \lambda- \frac{(m )^2}{1 - u^2} \right] S_{lm} = 0, \quad u=\cos\theta
\end{align}
where $\bar{K}_0=(r_{+}^2+a^2)\omega- m a$ and $\lambda=l(l+1)$.\\

Quasinormal modes are obtained by imposing the ingoing boundary condition at the horizon and the outgoing boundary condition at the spatial infinity.
 The QNM of the near-extremal Kerr-Newman black hole has been computed analytically in \cite{Hod:2008se}. The procedure to compute is the same which we describe in section \eqref{RN}. Since we describe all the steps already in section \eqref{RN}, we briefly mention the procedure to obtain the solution and write the expression of the quasinormal modes.
 
We begin by defining a new set of variables,
$$z = \frac{r - r_+}{r_+ - r_-} ; \quad \hat{\omega} = \frac{2(\omega - m\Omega)(r_{+}^2+a^2)}{r_{+}-r_{-}} ; \quad k \equiv \omega(r_+ - r_-) ,$$

Under this redefinition, one obtains a hypergeometric equation and the solution which satisfies the ingoing boundary condition near the horizon $r\rightarrow r_+$ with $k\,z \ll 1$ is given by

\begin{align}\label{nKN}
    f(z)= z^{ - i\hat{\omega}} (z + 1)^{ + i\hat{\omega}} \, {}_2F_1(-l , l + 1; 1  - 2i\hat{\omega}; -z)
\end{align}

In the asymptotic limit, $z\gg M$, the solution of the wave equation can also be obtained
\begin{align}\label{asKN}
  f(z)|_{z\gg M}&\sim  C_1 e^{-ikz} z^{l } {}_1F_1(l  + 1; 2l + 2; 2i \,k\, z) \nonumber\\
  &\quad\quad\quad\quad+ C_2 e^{-ikz} z^{1 - l  - 1} {}_1F_1(-l ; -2l; 2i\,k\,z)
\end{align}

The coefficient $C_1$ and $C_2$ is determined by matching two solutions in the overlapping region $|\hat{\omega}|+1\ll z\ll \frac{1}{k} $ 
\begin{align}
    C_1 &= \frac{\Gamma(2l + 1)\Gamma(1 - 2i\hat{\omega})}{\Gamma(l + 1)\Gamma(l + 1 - 2i\hat{\omega})}\\
    C_2&=\frac{\Gamma(-2l - 1)\Gamma(1  - 2i\hat{\omega})}{\Gamma(-l )\Gamma(-l - 2i\hat{\omega})} 
\end{align}
Once, we evaluate $C_1$ and $C_2$, we can plug these constants into the asymptotic solution \eqref{asKN} and take $z \rightarrow \infty$. We demand the outgoing part of the solution should vanish when $z\rightarrow \infty$.
\begin{comment}
    This equation can be solved easily in Mathematica.
\begin{align}
    f(z)&=C_1 z^{i\frac{\chi}{2}}(1-z)^{l+1+s}\, _2F_1(l+s+1,l+i \chi +1;s+i \chi +1;z)\nonumber\\
    &+C_2(1-z)^{l+s+1} z^{-s-\frac{i \chi }{2}} \, _2F_1(l-s+1,l-i \chi +1;-s-i \chi +1;z)
\end{align}
We now impose the ingoing boundary condition at $z=0$ which implies $C_2=0$.

Let us now expand the solution with ingoing boundary condition around $z=1$ and obtain
\begin{align}
    \lim_{z\rightarrow 1}f(z)&=\frac{\pi  c_1 (z-1)^l \csc (2 \pi  l) (1-z)^{s-2 l} \Gamma (s+i \chi +1)}{\Gamma (-2 l) \Gamma (l+s+1) \Gamma (l+i \chi +1)}+\cdots
\end{align}
We can obtain the quasinormal modes by imposing vanishing Dirichlet boundary condition at $z=1$ (asymptotic boundary) which implies
\end{comment}
This leads us to the condition, $\Gamma(l-2i\hat{\omega}+1)=0$ which implies
\begin{align}
    l-2i\hat{\omega}+1=-n, \quad\quad n=0,1,2,\cdots
\end{align}
Therefore, quasinormal modes are given by
\begin{align}\label{qm}
\omega
&=m \Omega - i \kappa_e (n + l + 1), \quad \kappa_e=\frac{r_+ - r_-}{2(r_{+}^2+a^2)} 
\end{align}
Let us now focus on slow rotating `near-extremal' limit \cite{Hod:2008se}. In this limit, $r_{+}\sim r_{-}$ which implies $M\gg (M^2-a^2-Q^2)^{\frac{1}{2}}$.

From the expression of the black hole temperature, we obtain
\begin{align}\label{tempkn}
    2\pi T&=\frac{r_{+}^2-a^2-Q^2}{2 r_{+}(r_{+}^2+a^2)}\nonumber\\
    &\approx \frac{ 2 M (M^2-a^2-Q^2)^{\frac{1}{2}}}{2 r_{+}(r_{+}^2+a^2)}\nonumber\\
    &=\frac{r_{+}-r_{-}}{2(r_{+}^2+a^2)}.
\end{align}
In the second line, we use the expression of $r_{+} $ and imposed $M\gg (M^2-a^2-Q^2)^{\frac{1}{2}}$. To obtain the third line, we use $r_{+}\approx M$. 

Therefore, in this `near-extremality' condition quasinormal modes are given by 
\begin{align}\label{ne}
    \omega_{e}&=m \Omega - i 2 \pi T (n + l + 1) .
\end{align}
The analytic expression of QNM of the near-extremal Kerr-Newman black hole has been reported before in \cite{Hod:2008se} and we observe that $\Omega \rightarrow 0$ limit, it matches with the QNM of near-extremal RN black hole in \eqref{QNMRN} with $\mu \rightarrow 0$ and under $\omega r_+\ll 1$ condition.
To derive, the quasinormal modes we consider only the positive momentum modes along the $\phi$ direction ($m>0$ case). Similarly, one gets another set of quasinormal modes with the opposite momentum ($m<0$ case) and also the zero momentum case. In order to obtain the  one-loop partition function, we will require a complete set of quasinormal modes.
\section{One-loop determinant in Kerr-Newman black hole}\label{sec7}
In this section, we compute the one-loop determinant of the minimally coupled scalar field in the near-extremal Kerr-Newman black hole from the DHS formula. In the previous section, we compute the quasinormal spectrum in the near-extremal limit of the KN black hole which we use to obtain the one-loop partition function in this section.
\subsection{One-loop determinant of a scalar field using quasinormal modes}
Since we are interested in evaluating the one-loop determinant in the near-extremal case, we will work with the quasinormal modes given in \eqref{ne}. But in principle, one can work with the quasinormal modes given in \eqref{qm} and at the end take the extremality limit. 

Let us first focus on the scalar field.
To compute the one-loop determinant for scalar field, we use the quasinormal modes in \eqref{ne} and the DHS formula \eqref{DHS}.
\begin{align}\label{KN}
    -\log\mathcal{Z}^{(1)}&=\sum_{p>0}\sum_{n,l,\geq 0}\sum_{m}\log\left(p+n+l+1+i\frac{m\Omega}{2\pi T}\right)+\log\left(p+n+l+1-i\frac{m\Omega}{2\pi T}\right)\nonumber\\
    &+\frac{1}{2}\sum_{n,l\geq0}\sum_{m}\log\left(n+l+1+i\frac{m\Omega}{2\pi T}\right)+\log\left(n+l+1-i\frac{m\Omega}{2\pi T}\right)\nonumber\\
    &=+\frac{1}{2}\sum_{k,l=0}^{\infty}\sum_{m}(2k+1)\log\left((k+l+1)^2+\frac{m^2\Omega^2}{(2\pi T)^2}\right)
\end{align}
Since we are concerned only about the logarithmic divergent piece, we stripped off the finite entire function which depends on $\Delta$.
In the second line, we combined the sum over $p$ and $n$ into a single sum over $k$.

To perform the sum over modes, we use the integral representation of logarithm.
\begin{align}\label{logiden}
    -\log \,y&=\int_0^{\infty}\frac{d\tau}{\tau}\left(e^{-y\tau}-e^{-\tau}\right)
\end{align}
Since the second term can be thought of as sum over degeneracy, by the same logic presented in section \eqref{RN}, we only consider the first term to compute the one-loop determinant \cite{Mukherjee:2021alj}.

We now proceed with the first term
\begin{align}
  -\log\mathcal{Z}^{(1)}&=
  \int_{\epsilon}^{\infty} \frac{d\tau}{2\tau} e^{-\frac{\epsilon^2}{4\tau}}\sum_{k,l=0}^{\infty}\sum_{m}(2k+1) e^{-\tau \left( \left( k+l+1 \right)^2 +\frac{m^2\Omega^2}{(2\pi T)^2}\right)}.  
\end{align}

We now use the Hubbard-Stratonovich trick to perform the sum over $k$.
\begin{align}
    \sum_{k=0}^{\infty} (2k+1) e^{-\tau \left( k+l+1\right)^2}
= \int_{C}\frac{du}{\sqrt{4\pi \tau}} e^{-\frac{u^2}{4\tau}} f_l(u).
\end{align}
where $C=\mathbb{R}+i\delta$ and $f_l(u)$ is given by
\begin{align}
    f_l(u) = \sum_{k=0}^{\infty} (2k+1)e^{iu(k+l+1)}
= \frac{e^{i (l+1) u}+e^{i (l+2) u}}{\left(-1+e^{i u}\right)^2}.
\end{align}
We can now perform the integral over $\tau$ 
\begin{align}
   -\log\mathcal{Z}^{(1)}&=\sum_{l\geq 0}\sum_{m=-\infty}^{\infty}\int_{C} \frac{du}{2\sqrt{u^2+\epsilon^2}}e^{-i \frac{| m\Omega|}{2\pi T} \sqrt{u^2+\epsilon^2} }f_l(u)
\end{align}
We now rotate the contour by substituting $u=it$ and sum over $m$ along the appropriate sides of the branch as shown in fig. \eqref{fig2}
\begin{align}
     -\log\mathcal{Z}^{(1)}&=\sum_{l\geq 0}\int_{\epsilon}^{\infty}\frac{dt}{2\sqrt{t^2-\epsilon^2}}\left(1+2\sum_{m=1}^{\infty}e^{-\frac{m\Omega}{2\pi T}\sqrt{t^2-\epsilon^2}}\right)f_{l}(u=it)\nonumber\\
     &=\int_0^{\infty}\frac{dt}{2t}\frac{e^t \left(e^t+1\right) \left(e^{\frac{t \Omega }{2 \pi  T}}+1\right)}{\left(e^t-1\right)^3 \left(e^{\frac{t \Omega }{2 \pi  T}}-1\right)}
\end{align}
In the last line, we perform a sum over $l$ and take $\epsilon \rightarrow 0$ limit.

It is now worth looking at the case $\Omega=2\pi T$. In this value of the angular velocity at the horizon, we obtain
\begin{align}\label{2sc}
     -\log\mathcal{Z}^{(1)}|_{\Omega=2\pi T}&=\int_0^{\infty}\frac{dt}{2t}\frac{1+e^{-t}}{1-e^{-t}}\frac{e^{-t}+e^{-2t}}{(1-e^{-t})^3}
\end{align}
Note that, at this value of the $\Omega$, the one-loop determinant of a scalar field in the near-extremal Kerr-Newman black hole reduces to the one-loop determinant of a scalar in the near-horizon extremal RN black hole with $AdS_2\times S^2$ geometry as given in \eqref{1sc}. Therefore, the logarithmic divergent piece also agrees with the near-horizon extremal RN case given in \eqref{f3}.
  \begin{figure}[h]
\centering
\begin{tikzpicture}[thick,scale=0.85]
\filldraw[magenta] 
                (0,0.5) circle[radius=3pt]
                (0,-0.5) circle[radius=3pt];
                \filldraw[orange] 
                (0.1,0) circle[radius=2pt]
                (0.5,0) circle[radius=2pt]
              (1,0) circle[radius=2pt]  
              (1.5,0) circle[radius=2pt]
              (2,0) circle[radius=2pt]
               (2.5,0) circle[radius=2pt]
                (3,0) circle[radius=2pt]
               (3.5,0) circle[radius=2pt]
                (4,0) circle[radius=2pt]
                (-0.1,0) circle[radius=2pt]
                (-0.5,0) circle[radius=2pt]
              (-1,0) circle[radius=2pt]  
              (-1.5,0) circle[radius=2pt]
              (-2,0) circle[radius=2pt]
               (-2.5,0) circle[radius=2pt]
                (-3,0) circle[radius=2pt]
               (-3.5,0) circle[radius=2pt]
                (-4,0) circle[radius=2pt] ;
\draw [decorate,decoration=snake] (0,-0.5) -- (0,-4);
\draw [decorate,decoration=snake] (0,0.5) -- (0,4);
\draw [postaction={decorate,decoration={markings , 
mark=at position 0.55 with {\arrow[black,line width=0.5mm]{<};}}}](0.2,1) arc[start angle=0, end angle=-180, radius=0.2cm];
\draw
[
postaction={decorate,decoration={markings , 
mark=at position 0.20 with {\arrow[red,line width=0.5mm]{>};}}}
][blue, thick] (-4,0.2)--(0,0.2);
\draw[blue, thick] (0,0.2)--(4,0.2);
\draw (0.4,1) node{$\mathbf{\epsilon}$};
\draw (0.4,-1) node{$\mathbf{-\epsilon}$};
\draw[gray, thick] (0,0) -- (0,4);
\draw[gray, thick] (0,0) -- (0,-4);
\draw[gray, thick] (0.2,4) -- (0.2,1);
\draw[gray, thick] (-0.2,1) -- (-0.2,4);
\draw[gray, thick] (0,0) -- (4,0);
\draw[gray, thick] (0,0) -- (-4,0);
\draw[gray, thick] (2,3) -- (2,3.4);
\draw[gray, thick] (2,3) -- (2.5,3);
\draw (2.3,3.3) node{$\mathbf{u}$};
\draw [red,thick](-2,0.5) node{$\mathbf{C}$};
\draw [gray,thick](0.5,3) node{$\mathbf{C'}$};
\end{tikzpicture}
\caption{The contour in the $u$-plane} \label{fig1}
\qquad
\centering
\begin{tikzpicture}[thick,scale=0.85]
\filldraw[magenta] 
                (0.5,0) circle[radius=3pt]
                (-0.5,0) circle[radius=3pt];
                \filldraw[orange] 
               ( 0,0.2) circle[radius=2pt]
                (0,0.5) circle[radius=2pt]
              (0,1) circle[radius=2pt]  
              (0,1.5) circle[radius=2pt] 
              (0,2) circle[radius=2pt] 
              (0,2.5) circle[radius=2pt]
                (0,3) circle[radius=2pt]
               (0,3.5) circle[radius=2pt]
                (0,4) circle[radius=2pt]
                (0,-0.1) circle[radius=2pt]
                (0,-0.5) circle[radius=2pt]
              (0,-1) circle[radius=2pt]  
              (0,-1.5) circle[radius=2pt]
              (0,-2) circle[radius=2pt]
               (0,-2.5) circle[radius=2pt]
                (0,-3) circle[radius=2pt]
               (0,-3.5) circle[radius=2pt]
                (0,-4) circle[radius=2pt] ;
\draw [decorate,decoration=snake] (-0.5,0) -- (-4,0);
\draw [decorate,decoration=snake] (0.5,0) -- (4,0);
\draw
[
postaction={decorate,decoration={markings , 
mark=at position 0.20 with {\arrow[red,line width=0.5mm]{>};}}}
][blue, thick] (0.5,0.2)--(4,0.2);
%\draw (7.9,0) node {$\Re(w)$};
%\draw (0,6.5) node {$\Im(w)$};
%\draw (5.5,2.5) node {$\Re(w)>0$};
%\draw (-5.5,2.5) node{$\Re(w)<0$};
%\draw (3,-3) node{$|w|=1$};
\draw (1,0.4) node{$\mathbf{\epsilon}$};
\draw (-1,0.4) node{$\mathbf{-\epsilon}$};
\draw[gray, thick] (0,0) -- (0,4);
\draw[gray, thick] (0,0) -- (0,-4);
\draw[gray, thick] (0,0) -- (4,0);
\draw[gray, thick] (-4,0) -- (0,0);
\draw[gray, thick] (2,3) -- (2,3.4);
\draw[gray, thick] (2,3) -- (2.5,3);
\draw (2.3,3.3) node{$\mathbf{t}$};
\end{tikzpicture}
\caption{The contour in the $t$-plane} \label{fig2}
\end{figure}

\begin{comment}
   % \subsection{Correction to the entropy, Shiraz's argument}
The  corresponding entropy
\begin{align}
    S_{\rm{BH}}&=\left(1-\beta\frac{\partial}{\partial\beta}\right)\log \mathcal{Z}|_{\beta=2\pi}.
\end{align}
But we can also write
\begin{align}
    \frac{\partial\log\mathcal{Z}}{\partial \beta}|_{\beta=2\pi}= \frac{\partial\log\mathcal{Z}}{\partial g_{\mu\nu}} \frac{\partial g_{\mu\nu}}{\partial \beta}|_{\beta=2\pi}.
\end{align}
Note that, all the $\beta$ variation comes from the background metric and we need to evaluate the change in free energy with the metric variation exactly at $\beta=2\pi$. But it is well known that for conformally flat space
\begin{align}
    \frac{\partial\log\mathcal{Z}}{\partial g_{\mu\nu}}=\frac{\sqrt{g}}{2}\langle T^{\mu\nu}\rangle.
\end{align}
Since $S^2\times AdS_2$ is Weyl equivalent to the flat space, the one-point function of the stress tensor vanishes.
\end{comment}

\section{One-loop determinant of the spin-1 field}\label{sec8}
Following the same principle, we can also compute the one-loop partition function for higher spin fields. To obtain the one-loop partition function for higher spins using the DHS formula, we require solving the quasinormal spectrum. In general, it may seem very hard, but fortunately, the radial and the angular wave equation for the spin-s field are known in the literature \cite{Dudley:1977zz, Berti:2005eb}. For the spin-s field, Dudley-Finley wave equations are also decomposed into radial and the angular part. 
\begin{equation}\label{angd}
\Psi_{slm}(t,r,\theta,\phi) = e^{im\phi} S_{slm}(\theta; a\omega)\, f_{slm}(r; a\omega)\, e^{-i\omega t},
\end{equation}
where  $\omega$ is the  frequency of the mode, 
$l$ is the quantum number associated with the angular momentum, and $m$ is the azimuthal quantum number with $-l \leq m \leq l$. 
\footnote{The angular and the radial part of the wave equations are also presented in \cite{Hod:2008se} (see equations (2) and (3) of \cite{Hod:2008se}).}
The angular part of the wave equation is given by
\begin{align}
\frac{1}{\sin \theta} \frac{\partial}{\partial \theta}\left(\sin \theta \frac{\partial S}{\partial \theta}\right)+\left[a^2 \omega^2 \cos ^2 \theta-2 a \omega s \cos \theta-\frac{(m+s \cos \theta)^2}{\sin ^2 \theta}+s+A\right] S=0 .
\end{align}
 The radial wave equation is given by
\begin{align}
\bar{\Delta}_H^{-s} \frac{d}{d r}\left(\bar{\Delta}_H^{s+1} \frac{d f_s}{d r}\right)+\left[\frac{\bar{K}_0^2-2 i s(r-M) \bar{K}_0}{\bar{\Delta}_H}-a^2 \omega^2+2 m a \omega-A+4 i s \omega r\right] f_s=0,
\end{align}
where $A=l(l+1)-s(s+1)+O(a\omega)$ is the separation constant between the radial and the angular wave equations.
To obtain the quasinormal modes, one solves the radial wave equation with the incoming boundary condition at the horizon.

The solution that satisfies the ingoing boundary condition is given by \cite{Hod:2008se}\footnote{For notational simplicity, we suppress the index $l$ and $m$ in the radial mode function $f_{slm}(r)$ and denote it simply by $f_s(r)$.}
\begin{align}\label{sol}
    f_s(z)&=z^{-s-i \hat{w}}(1+z)^{-s+i \hat{\omega}} \, {}_2F_1(-l -s, l-s + 1; 1 -s - 2i\hat{\omega}; -z),\quad\hat{\omega}=\frac{2(\omega - m\Omega)(r_{+}^2+a^2)}{r_{+}-r_{-}}
\end{align}
Following the procedure outlined in Section~\ref{DF}, the quasinormal modes of a slowly rotating near-extremal Kerr–Newman black hole yield the same expression for higher-spin fields as well~\cite{Hod:2008se}.

\begin{align}
    \omega_* = m \Omega - i \frac{r_+ - r_-}{2(r_{+}^2+a^2)} (n + l + 1) ,\quad m>0.
\end{align}

Since we have the same expression of QNM for higher spin perturbation, one would naively expect that the contribution to the one-loop determinant is also the same as scalar perturbation. However, we will show that the modes $p=-(s-1),\cdots (s-1)$ are not regular at the horizon, and therefore one has to remove these modes to compute the one-loop determinant.

Let us now expand the ingoing modes \eqref{sol} near the horizon $z\rightarrow 0$,
\begin{align}
    \lim_{z\rightarrow 0}f_s(z)=z^{-s-i \hat{w}} \left(1+\left(\frac{i l (l+1)}{2 \hat{w}}-s+i \hat{w}\right) z+O\left(z^2\right)\right)
\end{align}

from the DHS's argument, we vary $\Delta$ or $m^2$ in such a way that the QNM satisfies
\begin{align}\label{cond}
    \frac{i \omega_*}{2\pi T}=-|p|.
\end{align}
We will investigate the mode functions at these values of the masses (or at the QNMs). We now take the near-extremal limit $r_{+}\sim r_{-}$ and using the expression of the temperature \eqref{tempkn}, we get
\begin{align}
    \hat{\omega}=\frac{\omega-m \Omega}{2\pi T}.
\end{align}
Now, when $\omega$ satisfies the QNM condition \eqref{cond}, the mode function goes as
\begin{align}
    \lim_{z\rightarrow 0} f_s(z)\sim z^{-s+|p|}\left(1+O(z)\right)
\end{align}
Therefore, the mode functions become irregular at the horizon when $s>|p|$. To obtain, the one-loop determinant, we remove such modes. 

 The irregular behavior of higher spins at the horizon has also been observed before \cite{Grewal:2022hlo} and the corresponding edge partition function has been constructed for the higher spins for the black holes in $AdS$.

Let us compute the one-loop determinant for the spin-1 field. In this case, $p=0$ mode is irregular at the horizon which we have to remove.

The $p=0$ mode corresponds to the second term of \eqref{KN},
\begin{align}
 -\log\mathcal{Z}|_{p=0}&=   \frac{1}{2}\sum_{n\geq0, l>0}\sum_{m}\log\left(n+l+1+i\frac{m\Omega}{2\pi T}\right)+\log\left(n+l+1-i\frac{m\Omega}{2\pi T}\right)
\end{align}
Following the same procedure of scalar one-loop determinant, we obtain
\begin{align}
    -\log\mathcal{Z}|_{p=0}&=\int_0^{\infty}\frac{dt}{2t}\sum_{n\geq 0,l>0}e^{-t(n+l+1)}\left(1+2\sum_{m=1}^{\infty}e^{-\frac{m \Omega}{2\pi T}}\right)\nonumber\\
    &=\int_0^{\infty}\frac{dt}{2t}\frac{1}{\left(e^t-1\right)^2}\left(\frac{2}{e^{\frac{t \Omega }{2 \pi  T}}-1}+1\right).
\end{align}
In the limit, $\Omega \rightarrow 2\pi T$, we compute
\begin{align}
  \lim_{\Omega\rightarrow 2\pi T}   -\log\mathcal{Z}|_{p=0}&=-\int_0^{\infty}\frac{dt}{2t}\frac{1}{\left(e^t-1\right)^2}\frac{1+e^{-t}}{1-e^{-t}}.
\end{align}
The logarithmic divergent piece can be evaluated by expanding the integrand around $t=0$ and collecting the $1/t$ coefficient. But the logarithm of the area horizon will be half of the $1/t$ coefficient, because to regulate the integral, the lower limit of the integral is put to be a dimensionless ratio of IR and UV cutoff $a/\epsilon$ and therefore the log of the area will be twice as $A\sim a^2$.

Finally, we obtain the log divergent term for the vector field
\begin{align}
    -\log\mathcal{Z}(s=1)&=2(-\frac{1}{180}-\frac{1}{6})\log\frac{a^2}{\epsilon}\nonumber\\
    &=-\frac{31}{90}\log \frac{a^2}{\epsilon}.
\end{align}
The overall factor $2$ comes from the helicity factor $g_s=2$ in $d=4$ dimension for the spin-1 field. This logarithmic divergent piece agrees with \cite{Sen:2012kpz}.
\section{Conclusion} In this paper, we compute one-loop determinant of scalar and vector field in the  near-horizon geometry of the extremal Reissner Nordstr\"{o}m black hole. The calculation is based on the prescription by Denef-Hartnoll-Sachdev \cite{Denef:2009kn} which expresses the one-loop determinant as an infinite product over quasinormal modes. The proof of this prescription relies on the analytic properties of the partition function. We can think of the one-loop determinant as a meromorphic function in the mass parameter, and this meromorphic function is completely determined from the location of the poles (up to an entire function). The pole structure occurs when we encounter a zero mode of the wave equation of the corresponding field, which should also obey the periodicity condition in the Euclidean time direction. 

To implement the DHS prescription, we first analytically obtain the quasinormal modes in the extremal black hole. The quasinormal modes are obtained by imposing the ingoing boundary condition at the horizon, and the outgoing boundary condition at the asymptotic infinity. To solve the wave equation, we first obtain the solution in the asymptotic regime $r \rightarrow\infty$ with the outgoing boundary condition. We then also obtain the solution in the near-horizon region under the condition $r_+ T\ll 1$ with the ingoing boundary condition. We match these two solutions in the intermediate region and obtain the quasinormal modes. From the quasinormal spectrum, we evaluate the one-loop determinant of a scalar field and extract the logarithmic divergent piece which agrees with \cite{Banerjee:2010qc}. We also compute the one-loop determinant in the Kerr-Newman black hole using the DHS prescription and observe that when the angular velocity at the horizon is tuned to a specific value $\Omega =2\pi T$, the one-loop determinant of a scalar field at a finite-temperature Kerr-Newman black hole reduces to the one-loop partition function of the zero-temperature extremal black hole. At this specific value of the angular velocity of the horizon, the relation between the partition function of the finite-temperature Kerr-Newman  black hole and the zero-temperature extremal black hole has been shown before \cite{,H:2023qko}. 

However, we have not computed the finite- temperature correction of the partition function in the near-extremal limit, which we will investigate in the near future. Perhaps, in the finite temperature case, one requires a first-order perturbative correction to the quasinormal modes. We hope that the perturbative correction to the leading order in the quasinormal modes will give us the finite temperature correction to the one-loop partition function.  We also hope to generalize our computation for the fermions and higher spin fields in future works. The equality of the one-loop partition functions between near-extremal Kerr-Newman black hole at a finite but small temperature and the near-horizon extremal black hole may also be possible to generalize for the finite temperature (may not be small) Kerr-Newman black hole. The difficulty will be to obtain the QNM spectrum which is hard to obtain analytically. Possibly some numerical techniques may help in this case. 

As a part of generalization, one can compute the one-loop partition function minimally coupled gravitons and higher derivative fields in the near-horizon extremal RN black hole. The one-loop partition function of conformal higher derivative spin fields on $S^a\times AdS_b$ has been evaluated in \cite{Mukherjee:2021alj,Mukherjee:2021rri} and the connection with the entanglement entropy has also been established. It will also be interesting to understand the  edge modes \cite{David:2022jfd,Mukherjee:2023ihb} and the connection with the Harish-Chandra character representation of the partition function \cite{David:2020mls,David:2021wrw} in the near-extremal black holes.

It will be good to understand and evaluate the quasinormal modes of non-extremal black holes analytically and, evaluate the one-loop determinant and check against \cite{Sen:2012dw}. Recently, there has also been some progress in the one-loop determinant calculation for Kerr black holes in $(A)dS$ space \cite{Arnaudo:2024rhv} analytically from quasinormal modes. It will be nice to generalize for the charged rotating black holes and one can try to adapt some analytic handle of the quasinormal modes \cite{Amado:2021erf,BarraganAmado:2023wxt}.

It would be very interesting to compute the one-loop corrections from the graviton modes in the near-extremal limit, taking into account the full geometry, which is expected to reproduce the characteristic scaling \(Z \sim T^{3/2}\), and to make a direct connection with the Euclidean analysis of~\cite{Kolanowski:2024zrq}. A necessary first step is to determine the appropriate analytic continuation of the Lorentzian mode functions so that they can be matched to their Euclidean counterparts, and then to identify the resulting modes responsible for the characteristic scaling of the partition function. However, carrying out such an analysis requires substantial additional development and lies beyond the scope of the present work. We hope to return to this direction in the future.

\acknowledgments

The author wishes to thank Justin David and Shiraz Minwalla for discussions and valuable inputs in this project. The author also thanks Indian Institute of Science Education and Research (IISER), Pune and the organizers of the conference “Future Perspectives on QFT and Strings ” for hospitality while this work was in progress.
\appendix 
\section{Justification of the infinite-$n$ sum}\label{jna}
 In the limit, $\mu\rightarrow 0$, the quasinormal modes in \eqref{QNMRN}, become
\begin{align}
    \omega_*=-i 2\pi T(n+l+1).
\end{align}
We have obtained the quasinormal modes in \eqref{QNMRN} 
under the condition 
\(|\omega_*| r_+ \ll 1\).
This condition naturally imposes a cutoff on the mode number at 
\begin{equation}
   ( n+l)_{\rm max}\,\sim\,\frac{1}{r_+ T}.
\end{equation}
For $l=0$ mode, the cut-off on the mode $n_{\rm{max}}\sim \frac{1}{r_+ T}$.
In this appendix we demonstrate that the contribution of modes with 
\(n \gtrsim n_{\rm max}\) to the one-loop determinant does not modify the logarithmic divergence appearing in \eqref{one-loopdetsc}.  
Since the cutoff is set at \(n_{\rm max}\sim 1/(r_+T)\), we introduce the same cutoff on the redefined mode number \(k'=n+p\), in the third line of \eqref{one-loopdetsc}, with the Matsubara modes $p> 0$, likewise restricted by this bound.  
We therefore consider the sum over modes with 
\begin{equation}
    k' \,\gtrsim\, \frac{1}{r_+ T}.
\end{equation}

Let us now examine the divergent structure of the one-loop determinant for $l=0$ mode, as written in the third line of \eqref{one-loopdetsc}:
\begin{equation}
    -\log Z^{(1)}_{l=0}
    \;\sim\; 
    \sum_{k'\,\gtrsim\,\frac{1}{r_+T}}^{\infty}
    (2k'+1)\,\log\!\Big[\left(k'+\tfrac{1}{2}\right)^2+\delta^2\Big].
\end{equation}
In the limit, $\mu\rightarrow 0$, we have $|\delta|=(l+\frac{1}{2})$.
Introducing $u=k'+\tfrac{1}{2}$, the prefactor simplifies as $(2k'+1)=2u$, so that
\begin{equation}
    -\log Z^{(1)}_{l=0}\;\sim\; 
    2\sum_{u\sim n_0}^{\infty} u\,\log(u^2+\delta^2),
    \qquad n_0\;\equiv\;\frac{1}{r_+T}+\frac{1}{2}.
\end{equation}
The sum can be approximated using the Euler--Maclaurin formula:
\begin{align}
    \sum_{u=n_0}^{\Lambda} f(u) 
    &= \int_{n_0}^{\Lambda} f(x)\,dx 
    + \tfrac{1}{2}\Big[f(n_0)+f(\Lambda)\Big] 
    + \cdots,
    \label{eq:EMformula}
\end{align}
where in our case $f(u)=2u\log(u^2+\delta^2)$.
The integral can be performed very easily:
\begin{align}
  I_{\rm{int}}=  \int u\log(u^2+\delta^2)\,du
    &= \frac{1}{2}\Big[(u^2+\delta^2)\log(u^2+\delta^2)-(u^2+\delta^2)\Big].
\end{align}
We use a cut-off $\Lambda$ as an upper limit of the sum, which we eventually take it to $\infty$. The integral contribution can be written as:
\begin{align}\label{intexp}
    I_{\rm int}(\Lambda,n_0) 
    &= \Big[(u^2+\delta^2)\log(u^2+\delta^2)-(u^2+\delta^2)\Big]_{u=n_0}^{u=\Lambda}.
\end{align}
For large $u$, $\log(u^2+\delta^2)=2\log u+\mathcal{O}(u^{-2})$, and hence the $n_0$-dependent part scales as
\begin{equation}
    I_{\rm int}\;\sim\;-\,2\,n_0^2\log n_0 \;+\; \mathcal{O}(n_0^2).
\end{equation}
Note that, in this case for $l=0$ mode, $|\delta|=\frac{1}{2}\ll n_0$, therefore, we can always expand in large $n_0$.
There are corrections to the leading order term and
the first correction term in~\eqref{eq:EMformula} is given by
\begin{equation}
    \tfrac{1}{2} f(n_0) \;\sim\; n_0\log n_0,
\end{equation}
while the higher-order divergent terms in the  Euler--Maclaurin series produce only inverse powers of $n_0$. Therefore, the subleading corrections involve $n_0\log n_0$ and polynomial divergent pieces but no pure logarithmic divergent contribution.

Collecting the leading order and sub-leading divergent terms, we get
\begin{equation}
    -\log Z^{(1)}_{l=0} \;\sim\;
    -\,2\,n_0^2\log n_0
   +\,n_0\log n_0
    \;+\; \cdots
\end{equation}
Since,
\begin{equation}
    n_0 \;\sim\; \frac{1}{r_+T},
\end{equation}
we conclude that the divergences are of the type $\frac{1}{r^2_+T^2}\log(r_+T)$, $\frac{1}{r_+T}\log(r_+T)$, and polynomial powers of $\frac{1}{r_+ T}$, but there is no isolated pure logarithmic divergent term present in the series. It is evident from the above analysis that, including the modes above $n\geq n_{\rm{max}}\sim \frac{1}{r_+T}$ does not alter the logarithmic divergent term.

The above analysis can be straightforwardly extended to each angular mode labeled by $l$, with the natural cutoff  
\begin{equation}
    (n+l)_{\rm max} \,\sim\, \frac{1}{r_+T}.
\end{equation}
For fixed $l\ll n_0$, and hence fixed $|\delta|$, the condition $\tfrac{|\delta|}{n_0}\ll 1$ continues to hold. 
In every case, as demonstrated, no purely logarithmic divergent term arises. 
It follows that the sum over $n$ can be extended all the way to infinity without modifying the logarithmic divergence. 
After performing the sum over $n$, the subsequent summation over all $l<n$-modes in the one-loop determinant similarly does not introduce any additional logarithmic divergence. 

 Similarly, for $l\gtrsim n_0$ modes, we can perform a similar analysis and show that there is no stand-alone log-divergent piece. Note that, for $l\gg n_0$, in the expression \eqref{one-loopdetsc}, we can sum over $l$ first in this case, 
 \begin{align}
     -\log Z^{(1)}&\sim \sum(2k'+1)\sum_{l\gtrsim n_0}(2l+1)\left[\log\Big[(l+\frac{1}{2})^2-(k'+\frac{1}{2})^2\Big]\pm i\pi\right]\nonumber\\
    &\sim \sum(2k'+1)\sum_{l\gtrsim n_0}(2l+1)\log\Big[(l+\frac{1}{2})^2+\delta'^2], \quad \delta'^2=-(k'+\frac{1}{2})^2
 \end{align}
 By the same logic presented as before, we would find no logarithmically divergent term (since $k' \sim n\sim O(1)$). Therefore, we can independently extend the sum over $l$ and keep $n\ll n_0$.

Let us now summarize. We can split the angular-momentum range into the three regimes (i) $l\ll n_0$, (ii) $l\sim n_0$, and (iii) $l\gg n_0$.  

 For (i), $|\delta|=l+\tfrac12\ll n_0$ and with $u=k'+\tfrac12$ the large-$u$ expansion
 $\log(u^2+\delta^2)=2\log u+O(u^{-2})$ is valid. Applying Euler-Maclaurin to
 $f(u)=2u\log(u^2+\delta^2)$ (first case) yields boundary and integral pieces scaling as
 $n_0^2\log n_0$, $n_0\log n_0$ and polynomial powers of $n_0$, but no isolated
 $\log$-divergent term. Thus the large-$n$ tail does not produce a pure logarithmic divergence.  

 For (iii), we have the same situation as (i), where  $l \gg n_0$ with $n \sim O(1)$ and hence there is no large-$l$ tail to generate a new $\log$. The borderline case (ii) does not produce an isolated $\log$ either — its contributions are of the same polynomial or $n^2_0\log n_0$ type.  

% Finally, when summing over $l$ and $n$ one must keep the regularization and the branch choice of the logarithm consistent (the analytic continuation that yields $\log[(l+\tfrac12)^2-(k'+\tfrac12)^2]\pm i\pi$ is acceptable provided the branch is stated). Under these conditions, the preceding analysis covers all modes and no additional isolated $\log$ term arises when extending the sums to infinity.

% Bibliography

%% [A] Recommended: using JHEP.bst file
%% \bibliographystyle{JHEP}
%% \bibliography{biblio.bib}

%% or
%% [B] Manual formatting (see below)
%% (i) We suggest to always provide author, title and journal data or doi:
%% in short all the informations that clearly identify a document.
%% (ii) please avoid comments such as "For a review'', "For some examples",
%% "and references therein" or move them in the text. In general,,, please lwave only references in the bibliography and move all
%% accessory text in footnotes.
%% (iii) Also, please have only one work for each \bibitem.

\bibliographystyle{JHEP}
\bibliography{reference.bib}
\end{document}